\documentclass[twocolumn, preprintnumbers, amsmath, amssymb]{revtex4-1}

\usepackage{graphicx}
\usepackage{dcolumn}
\usepackage{bm}
\usepackage{longtable}

\def\bea {\begin{eqnarray}}
\def\eea {\end{eqnarray}}

\def\be {\begin{equation}}
\def\ee {\end{equation}}

\begin{document}


\title{First Observation of Multiple Transverse Wobbling Bands of Different Kinds in $^{183}$Au}

\author{S. Nandi$^{1,2}$,
G. Mukherjee$^{1,2}$\thanks{\email gopal@vecc.gov.in},
Q. B. Chen$^{3}$,
S. Frauendorf$^{4}$,
R. Banik$^{1,2}$,
Soumik Bhattacharya$^{1,2}$,
Shabir Dar$^{1,2}$,
S. Bhattacharyya$^{1,2}$,
C. Bhattacharya$^{1,2}$,
S. Chatterjee$^5$,
S. Das$^5$,
S. Samanta$^5$,
R. Raut$^5$,
S.S. Ghugre$^5$,
S. Rajbanshi$^{6}$,
Sajad Ali$^{7}$,
\altaffiliation[Present address: ]
{Govt. Degree College}
H. Pai$^{8}$,
Md. A. Asgar$^{9}$,
S. Das Gupta$^{10}$,
P. Chowdhury$^{11}$,
A. Goswami$^{8}$\thanks{deceased}
}
\affiliation{$^{1}$Variable Energy Cyclotron Centre, 1/AF, Bidhan Nagar, Kolkata - 700064, India}
\affiliation{$^{2}$Homi Bhabha National Institute, Training School Complex, Anushakti Nagar, Mumnai - 400094, India}
\affiliation{$^{3}$Physik-Department, Technische Universit\"{a}t M\"{u}nchen, D-85747 Garching, Germany}
\affiliation{$^{4}$Department of Physics, University of Notre Dame, Notre Dame, Indiana 46556, USA}
\affiliation{$^{5}$UGC-DAE CSR, Kolkata Centre, Kolkata 700098, India}
\affiliation{$^{6}$Department of Physics, Presidency University, Kolkata 700043, India}
\affiliation{$^{7}$Government General Degree College at Pedong, Kalimpong 734311, India}
\affiliation{$^{8}$Saha Institute of Nuclear Physics, Kolkata 700064, India}
\affiliation{$^{9}$Department of Physics, Prabhat Kumar College, Contai 721404, India}
\affiliation{$^{10}$Victoria Institution (College), Kolkata 700009, India}
\affiliation{$^{11}$University of Massachusetts Lowell, Lowell MA 01854, USA}

\date{\today}

\begin{abstract}

We report the first observation of two wobbling bands in $^{183}$Au,
both of which were interpreted as the transverse wobbling (TW) band
but with different behavior of their wobbling energies as a function
of spin. It increases (decreases) with spin for the positive
(negative) parity configuration. The crucial evidence for the wobbling 
nature of the bands, dominance of the $E2$ component in the $\Delta I = 1$ 
transitions between the partner bands, is provided by the simultaneous 
measurements of directional correlation from the oriented states (DCO) 
ratio and the linear polarization of the $\gamma$ rays.
Particle rotor model calculations with triaxial deformation
reproduce the experimental data well. A value of spin, $I_m$, has been
determined for the observed TW bands below which the wobbling energy increases
and above which it decreases with spin. The nucleus $^{183}$Au is, so far,
the only nucleus in which both the increasing and the decreasing parts are
observed and thus gives the experimental evidence of the complete
transverse wobbling phenomenon.
\end{abstract}


\maketitle

Nuclear wobbling excitation is a manifestation of non-axial nuclear shape which
was first discussed by Bohr and Mottelson \cite{bohr}. The non-axial (triaxial)
nuclear shape appears due to the unequal nuclear mass distribution along the
three principal axes and implies three unequal moments of inertia about the
three principal axes. A triaxially deformed nucleus always tries to rotate
around the medium ($m$) axis having the largest moment of inertia but the presence
of the rotations around the other two axes i.e., short ($s$) and long ($l$), generates a
precession of the medium axis rotation about the space-fixed angular momentum axis,
similar to the classical wobbling motion of an asymmetric top \cite{landau}.
The energy spectrum of this excitation is given by \cite{bohr}:
$$
E = E_\textrm{rot} + (n_w+1/2)\hbar\omega_\textrm{wob}
$$
where, the term $E_\textrm{rot}$, corresponds to the rotation about the medium axis
while $n_w$ is the wobbling quanta and $\hbar\omega_\textrm{wob}$ is the wobbling
frequency with wobbling energy $E_\textrm{wob}$ = $\hbar\omega_\textrm{wob}$. This
generates a series of rotational bands with different n$_w$.

\begin{figure*}
\begin{center}
\includegraphics*[height=10.0cm, width=17.4cm, angle =0]{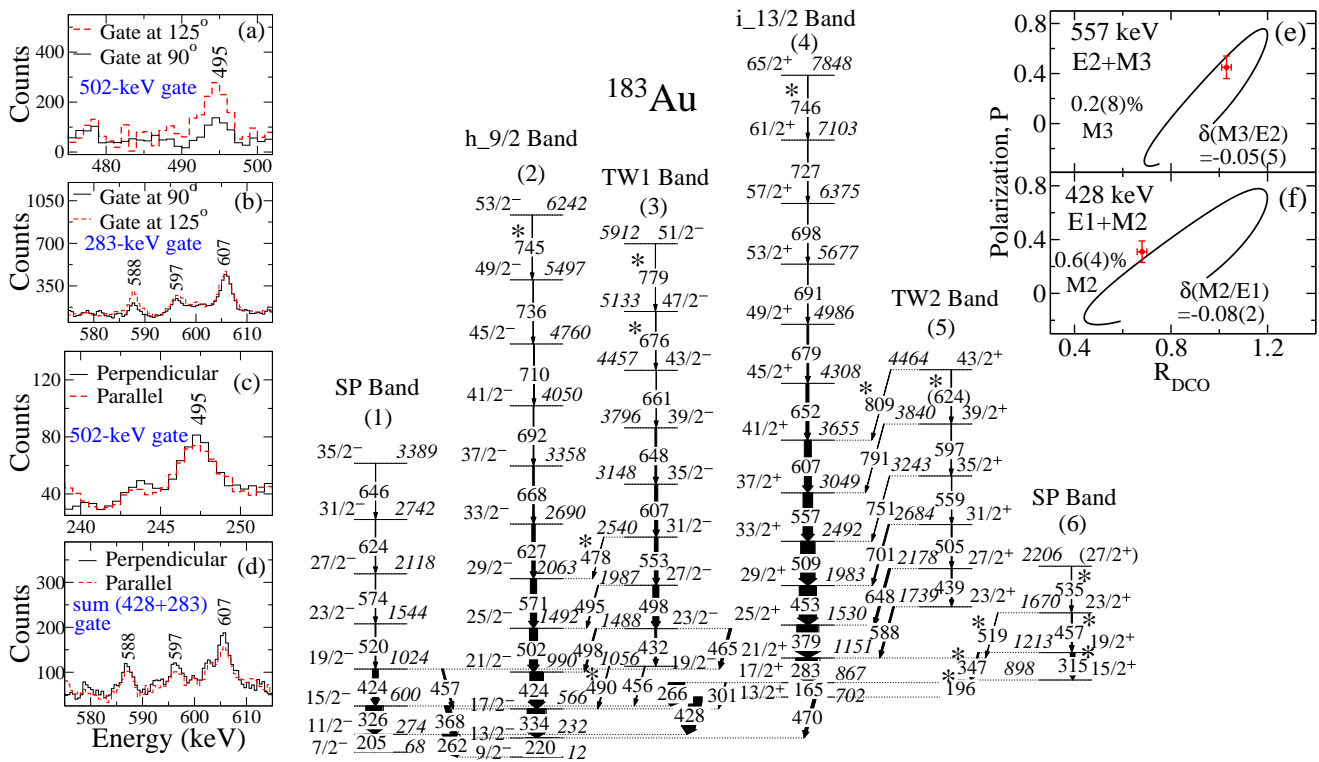}
\caption{\label{levelscheme} Level scheme of $^{183}$Au, established from the 
present work. Line widths are proportional to their intensities. The level 
energies are obtained by fitting the $\gamma$-ray energies using the code 
$GTOL$ \cite{gtol}. Representative gated $\gamma$-ray spectra for a few 
transitions, projected from the 2-dimensional DCO and polarization matrices, 
are shown in the insets (a), (b) (c) and (d). The spectra in (a) and (b) are 
used for DCO ratio (R$_{DCO}$) whereas (c) and (d) are used for linear 
polarization (P) measurements. The insets (e) and (f) show the experimental 
(symbol) and calculated (solid line) values of R$_{DCO}$ and P of two of the 
transitions in $^{183}$Au. The 557 keV is a known $E2$ transition decaying 
from 37/2$^+$ to 33/2$^+$ in band (4) and 428 keV is a known $E1$ transition
from 13/2$^+$ to 11/2$^-$ between band (4) and band (1). These transitions show 
very small mixing ratios ($\delta$) as they should be. 
The new transitions in the level scheme are marked by asterisks.
}

\end{center}
\end{figure*}

This exotic excitation has been observed only in a few odd-$A$ nuclei
\cite{163lu1,163lu2,161lu,165lu,167lu,167ta,105pd,133la,135pr,187au, NS2}.
In case of the odd-$A$ nuclei, the odd particle in high-j orbital couples with a 
triaxial core and modifies the wobbling motion. Depending on the coupling of the 
odd particle, two types of wobbling bands can be observed: Longitudinal Wobbling 
(LW) and Transverse Wobbling (TW) \cite{fra}. In LW, the angular momentum of the
odd particle aligns along the medium axis while in TW, it aligns along one of the 
perpendicular axes (short or long).

An extensive theoretical description of the wobbling motion has been given by 
Frauendorf and D\"onau \cite{fra} in terms of a quasiparticle triaxial rotor (QTR) 
model. Analytical expression for $\hbar\omega_\textrm{wob}$ has been derived with 
the assumption of ``Frozen Alignment"
and harmonic oscillation (HFA). It was shown that $E_\textrm{wob}$ increases as a
function of angular momentum ($I$) in case of LW which has been recently observed
experimentally in $^{133}$La \cite{133la} and $^{187}$Au \cite{187au}.
However, in case of TW, the variation of $E_\textrm{wob}$ is highly dependent on 
the values of the moments of inertia, $\mathcal{J}_{m}$, $\mathcal{J}_{s}$ and 
$\mathcal{J}_{l}$ along the medium, short and long axes, respectively, of the triaxial 
core. In general, $E_{\textrm{wob}}$ decreases with $I$. But in a situation where 
$\mathcal{J}_{m}$ is slightly larger than $\mathcal{J}_{s}$ and both are much larger 
than $\mathcal{J}_{l}$, the value of $E_\textrm{wob}$ increases
with the increase of $I$ for the lower values of $I$ and then decreases with $I$ \cite{fra}.
So far, in all the cases, in which the experimental evidence of TW bands have been reported,
only the decreasing part of the $E_\textrm{wob}$ has been observed. Therefore, it was generally
believed that the LW and TW can be distinguished from the variation of $E_\textrm{wob}$ with $I$;
increasing (decreasing) of which corresponds to LW (TW).

In this letter we report the observation of two TW bands in $^{183}$Au.
$E_\textrm{wob}$ has been found to increase with $I$ in one of these bands, the first example
of such a behavior of a TW band, while in the other band $E_\textrm{wob}$ decreases with $I$.
Thereby, it constitutes the only example of experimental observation of a complete scenario of
the transverse wobbling phenomenon.

The excited $n_w = 1$ wobbling bands decay to the $n_w$ = 0 yrast bands via $\Delta I =1$, 
collectively enhanced $E2$ transitions \cite{fra,187au}. On the other hand, in case of the 
signature partner (SP) bands (which occurs when the odd-particle angular momentum is not 
fully aligned with the rotation axis), the connecting transitions are predominantly 
$\Delta I = 1$, $M1$ type \cite{187au}. Therefore, experimentally the wobbling partner 
band and the SP band can be distinguished from
the nature of the connecting transitions between the partner band and the yrast band.
For clear identification
of wobbling, it is important to experimentally observe both the signature partner band and
the wobbling partner band decaying to the main band.

The excited states of $^{183}$Au in this work were populated using the
$^{169}$Tm($^{20}$Ne, 6n)$^{183}$Au fusion evaporation reaction at the bombarding
energy of 146-MeV from the K-130 cyclotron facility at the Variable Energy Cyclotron
Centre (VECC), Kolkata. A thick (23~mg/cm$^2$) $^{169}$Tm foil was used as the
target for this experiment. The $\gamma$ rays emitted from the residual nuclei
were detected by the Indian National Gamma Array ({\bfseries\small INGA}) facility
which was composed of eight Compton-suppressed (BGO anti-Compton shields) clover HPGe detectors
and two HPGe planar LEPS (Low Energy Photon Spectrometer). A total of $1.5 \times 10^{9}$
two- and higher-fold coincidence data with time stamp were recorded in a fast (250 MHz) digital
data acquisition system based on Pixie-16 modules of XIA~\cite{das}.
The data were sorted using the IUCPIX package \cite{das}, and were analyzed using
the radware package \cite{rad}. Several $E_{\gamma}$-$E_{\gamma}$ coincidence
matrices and a $E_{\gamma}$-$E_{\gamma}$-$E_{\gamma}$ cube were constructed
from which single- and double-gated spectra were projected for the analysis.

A new and improved level scheme of $^{183}$Au has been established (Fig.~\ref{levelscheme})
with 14 new $\gamma$ rays including a new band, the band (6), compared to the previous
studies \cite{183au_1, 183au_2}.
Representative gated spectra are shown in the insets in Fig.~\ref{levelscheme}.

For unambiguous assignment of multipolarity ($\lambda$) and type ($E/M$) of a $\gamma$-ray
transition as well as its mixing ratio $\delta$, in case of a mixed transition (e.g. $M1+E2$),
the combined measurements of linear polarization (P) and DCO (Directional Correlation from the
Oriented states) ratio ($R_{DCO}$), as described in Ref.~\cite{dco,pol1,pol2,rb,sn}, have been
performed.
The measured values of P and R$_{DCO}$ of a $\gamma$ ray have been compared with the values
calculated for several $\delta$. For a pure transition, the measured values would lie
close to the ones calculated with $\delta \sim 0$. This is demonstrated for
two transitions in $^{183}$Au, known to be E2 and E1 (see insets (b) and (c) of
Fig.~\ref{levelscheme}). In the case of $\Delta I =1$ transition with
enhanced $E2$ component, measured values would lie close to the calculated
ones with large $\delta$.

\begin{figure}
\begin{center}
\includegraphics*[height=8.0cm, width=8.5cm, angle =0]{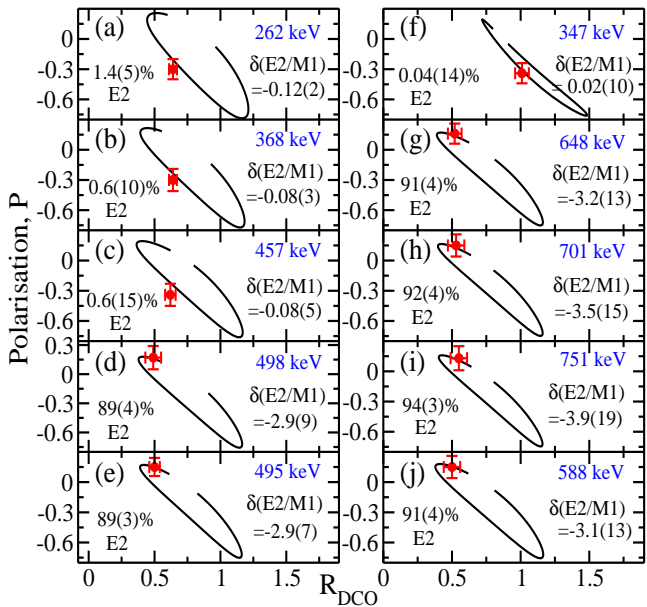}
\caption{Experimental (symbol) and calculated (solid line) values (for different mixing
ratios $\delta$) of DCO ratios (R$_{DCO}$) and linear polarization (P) of the connecting
transitions that decay to the negative parity band (2) (a $-$ e) and to the positive parity
band (4) (f $-$ j).
}
\label{dco_pol}
\end{center}
\end{figure}

Both negative- and positive-parity yrast bands, band (2) and (4), in $^{183}$Au have two side bands
each and are connected by $\Delta I = 1$ transitions. The measured and the calculated values of P
and R$_{DCO}$ for these connecting transitions are plotted in Fig.~\ref{dco_pol}.
The low values of $\delta$, obtained for the connecting transitions from bands (1) and (6) to the
respective negative and positive parity yrast bands, suggest that they are mostly of pure $M1$
type (about 1\% $E2$). On the other hand, quite large values of $\delta$ are resulted for the
transitions connecting the side bands (3) and (5) to the respective yrast bands, It clearly
indicates the predominantly $E2$ nature (about 90\% $E2$) of these transitions. Therefore,
the bands (1) and (6) have been identified as the SP bands, while the bands (3) and (5) are
identified as the one-phonon wobbling (n$_\omega$ = 1) bands. It is the first instance that
a pair of wobbling bands with positive and negative parity has been identified in a nucleus.

The positive and the negative parity bands in $^{183}$Au were assigned the configurations
$\pi i_{13/2}$ and $\pi h_{9/2}$, respectively \cite{183au_2}. Both of these orbitals
are situated above the $Z=82$ shell closure and, hence, they are of particle character with
large contribution from the low-$\Omega$ ($\Omega$ = projection of the particle angular
momentum on to the symmetry axis) Nilsson configuration. The quasiparticle aligned angular
momentum $i_x \approx 6.5 \hbar$, estimated for the positive parity yrast band from its level
energies, indicates fully aligned $\Omega = 1/2$, $i_{13/2}$ configuration. In case
of the negative parity band,  the estimated value of $i_x$ ($\approx 3.5 \hbar$) is somewhat
less than the fully aligned value ($= 4.5 \hbar$), suggesting a small mixing with larger
$\Omega$ orbitals. However, both of these situations favor the alignment of the odd particle
perpendicular to the medium axis and, hence, the occurrence of transverse wobbling.

It was shown in Ref.~\cite{fra} that the variation of the wobbling energy with spin
($I$) depends on the type (LW or TW) of wobbling motion.
Experimentally, the wobbling energy $E_\textrm{wob} = \hbar\omega_\textrm{wob}$ can be obtained from the
energy differences between the $n_w=1$ wobbling partner band and $n_w=0$ yrast band using the
relation \cite{105pd,133la,135pr,187au}
\noindent
$$
\hskip -40mm E_\textrm{wob} = E (I, n_w=1) -
$$
\vskip -9mm
$$
\hskip 15mm [E(I-1, n_w=0) + E(I+1, n_w=0)]/2,
$$
where $E(I)$ is the excitation energy of the state with angular momentum $I$.

\begin{figure}
\begin{center}
\includegraphics*[width=8 cm]{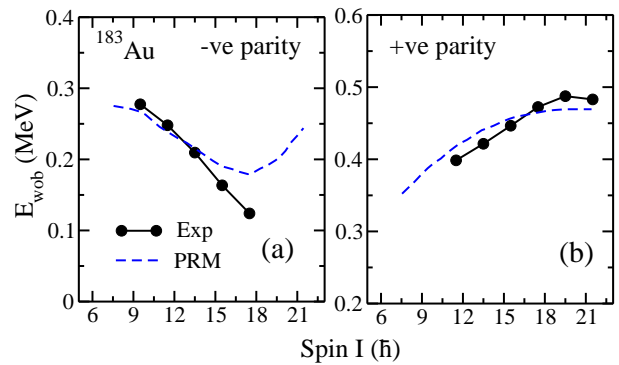}
\caption{Experimental wobbling energy $E_{\textrm{wob}}$ as a function of angular momentum $I$
for the negative parity (a) and the positive parity (b) bands in $^{183}$Au. The
theoretically calculated values are also shown. The error bars on experimental values are within the
size of the data points.
}
\label{Ewobb}
\end{center}
\end{figure}

The experimental values of $E_\textrm{wob}$ for both the wobbling bands,
identified in $^{183}$Au in the present work, are plotted in Fig.~\ref{Ewobb}.
It can be seen that the wobbling energy decreases (increases) with $I$
for the negative (positive) parity band and hence, it may be, in general, considered as
TW (LW). However, the low-$\Omega$, $i_{13/2}$ configuration for the
positive parity band is in contradiction to the LW geometry of the coupling of the odd
quasiparticle. Therefore, this band is, most likely, the initial part of a TW band,
predicted in Ref.\cite{fra} for the same $\pi i_{13/2}$ configuration in
$^{163}$Lu but has not been observed in any nucleus prior to this work.

In order to understand the observation of wobbling bands in
$^{183}$Au, theoretical calculations in the frame work of particle
rotor model (PRM)~\cite{Hamamoto2002PRC, fra, W.X.Shi2015CPC,
Streck2018PRC, Q.B.Chen2019PRC} has been performed. 
As the input to the calculations, the deformation parameters of 
$\beta = 0.30$ and $\gamma = 20.0^\circ$ and the moments of inertia
$\mathcal{J}_{m,s,l}=36.85, 25.70, 5.45~\hbar^2/\textrm{MeV}$ for
the negative parity band, while $\beta = 0.29$, $\gamma =
21.4^\circ$, and $\mathcal{J}_{m,s,l}=50.00, 37.52,
2.38~\hbar^2/\textrm{MeV}$ for the positive parity band have been
used. The deformation parameters are obtained by the covariant
density functional theory (CDFT)~\cite{mxd1, J.Meng2016book} with
PC-PK1 effective interaction~\cite{P.W.Zhao2010PRC}, and the moments
of inertia are fitted to the energy spectra. In both calculations,
the pairing gap $\Delta=12/\sqrt{A}=0.89~\textrm{MeV}$ is adopted.
It should be noted that the calculations can reproduce the
experimental $B(E2)$ values~\cite{183au_pj}, which justifies the
correct prediction of the deformation parameters by CDFT.
We have also carried out calculations by including cranking terms. 
The equilibrium deformations for both negative and positive parity 
bands change only 
within 0.01 for $\beta$ and $1^\circ$ for $\gamma$ from the bandhead 
to the highest spin, which justifies the assumption of a constant 
deformation for the PRM calculations.

\begin{figure}
\begin{center}
\includegraphics*[width=8.0 cm]{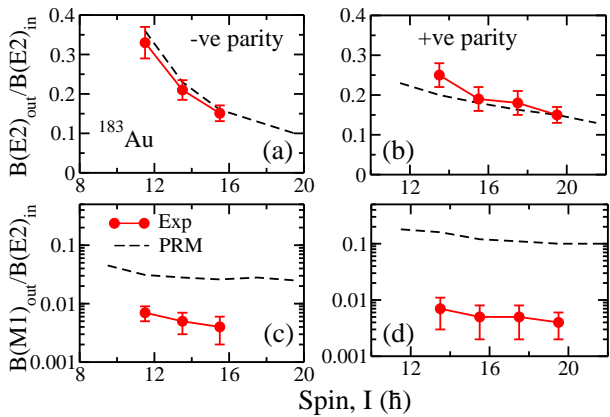}
\caption{Measured values of the ratio of transition probabilities,
$B(E2)_{\textrm{out}}/B(E2)_{\textrm{in}}$ and $B(M1)_{\textrm{out}}/B(E2)_{\textrm{in}}$,
determined from the $\gamma$-ray intensities, as a function of angular momentum $I$ for
the negative parity (a and c) and the positive parity (b and d) bands in $^{183}$Au.
The theoretical values calculated from PRM are also shown.}
\label{em}
\end{center}
\end{figure}

The results of the calculations are shown in Figs.~\ref{Ewobb} and 
\ref{em}. The wobbling energies ($E_{\textrm{wob}}$) for both the 
bands are well reproduced in the calculations. Moreover, the experimental
$B(E2)_{\textrm{out}}/B(E2)_{\textrm{in}}$ values, determined from the 
measured $\gamma$-ray intensities, agree well with the calculated ones 
(Fig.~\ref{em}) which justifies the correct inputs of triaxiality by
CDFT. 
The overestimation of the $B(M1)_{\textrm{out}}/B(E2)_{\textrm{in}}$ values
is attributed to the absence of scissors mode in the calculations \cite{Frauendorf2015PRC}.
The large $B(E2)_{\textrm{out}}/
B(E2)_{\textrm{in}}$ and small $B(M1)_{\textrm{out}}/B(E2)_{\textrm{in}}$ values further
support the wobbling interpretation for both the bands.

\begin{table}
\begin{center}
\caption{The moments of inertia along medium ($\mathcal{J}_m$), short
($\mathcal{J}_s$) and long ($\mathcal{J}_l$) axes obtained for the wobbling
bands in $^{183}$Au, $^{135}$Pr and $^{105}$Pd. The values for the later
two nuclei are taken from Ref.~\cite{fra} and \cite{105pd}, respectively.
The estimated values of $I_m$ based on HFA approximation are also given.
}
\label{tab1}
\begin{tabular}{ccccccc}
\hline
 & $^{183}$Au  & $^{183}$Au & $^{135}$Pr & $^{105}$Pd \\
 & $\pi i_{13/2}$ band & $\pi h_{9/2}$ band & $\pi h_{11/2}$ band & $\nu h_{11/2}$ band\\
\hline
$\mathcal{J}_m$ & 50.00 & 36.85 & 21.0 & 9.24 \\
$\mathcal{J}_s$ & 37.52 & 25.70 & 13.0 & 5.87 \\
$\mathcal{J}_l$ & 2.38  & 5.45  & 4.0  & 1.99 \\
$\mathcal{J}_m$/$\mathcal{J}_s$ & 1.33  & 1.43  & 1.61 & 1.57 \\
$I_m$ ($\hbar$) & 16.5  & 7.5  & 5.5 & 6.5 \\
\hline
\end{tabular}
\end{center}
\end{table}

The values of the three moments of inertia obtained for the two wobbling bands in $^{183}$Au
indicate that the ratios of both $\mathcal{J}_m / \mathcal{J}_l$ and
$\mathcal{J}_s / \mathcal{J}_l$ are much higher in case of the positive parity $\pi i_{13/2}$
band. It makes this band an ideal candidate for the observation of the, hitherto unobserved,
initial part of a TW band in which $E_\textrm{wob}$ increases with $I$ until a certain
value $I_m$ after which, it starts to decrease. The value of $I_m$ must be sufficiently
large in order to experimentally observe the increasing part. According to the harmonic
wobbling model with frozen orbital approximation (HFA) of Ref.~\cite{fra},
the larger is the value of the quasi particle angular momentum $j$, and closer is the value
of $\mathcal{J}_m$ / $\mathcal{J}_s$ to unity, the larger will be the value of $I_m$.

\begin{figure}
\begin{center}
\includegraphics*[width=8.0 cm]{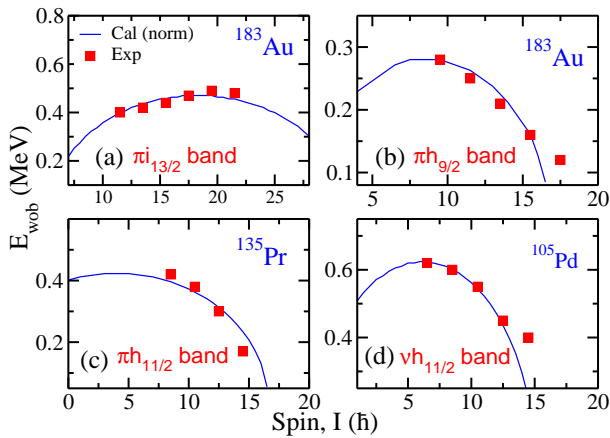}
\caption{Experimental and calculated values (see text for details) of wobbling energies as
a function of angular momentum ($I$) for the (a) positive ($i_{13/2}$) and (b) negative
($h_{9/2}$) parity wobbling bands in $^{183}$Au. For comparison, the same for the normal
deformed TW bands in $^{135}$Pr (c) and $^{105}$Pd (d) are also shown. Data for the
later two nuclei are obtained from Ref.\cite{135pr} and \cite{105pd}, respectively.
}
\label{ewob}
\end{center}
\end{figure}

The $I_m$ values estimated from the HFA model in $^{183}$Au have been compared,
in Table~\ref{tab1}, with the TW bands reported in the two other normal deformed nuclei,
$^{105}$Pd \cite{105pd} and $^{135}$Pr \cite{135pr}.
The configurations of the TW bands in $^{105}$Pd and $^{135}$Pr are $\nu h_{11/2}$
and $\pi h_{11/2}$, respectively. The value of $I_m$ is the largest for the $\pi i_{13/2}$
band in $^{183}$Au among these nuclei and is much more than the initial spin of its wobbling
band. Therefore, $E_\textrm{wob}$ for this band would fall in the increasing part of the predicted
TW band. This is further established
from Fig.~\ref{ewob} in which $E_\textrm{wob}$ calculated from the HFA model using the
fitted moments of inertia values $\mathcal{J}_{m,s,l}$ from Table~\ref{tab1}
are plotted as a function of spin along with the measured $E_\textrm{wob}$. The calculated
values are suitably normalised for comparison. Excellent agreement between experiment
and theory has been achieved for all the cases. It can be seen that the data points
corresponding to the $\pi i_{13/2}$ band in $^{183}$Au match with the increasing part
while for the others they fall nicely on the decreasing part of the curves for which the
values of $I_m$ are lower
than the lowest spin of the observed wobbling bands.

This shows, for the first time, an experimental evidence of
the general nature of the TW bands in terms of the variation of the wobbling energy
in which both increasing and decreasing parts of the $E_\textrm{wob}$ are observed.

It may also be pointed out that similar to the multiple chiral doublet (M$\chi$D)
bands \cite{mxd1}, the observation of the presence of multiple TW bands is also
an evidence of triaxial shape coexistence in $^{183}$Au. Though the triaxial
shape coexistence has been experimentally confirmed in several nuclei
through M$\chi$D bands (see for example \cite{mxd2,195tl} and references there in),
but its realization through the observation of multiple wobbling bands is being
reported for the first time.

In summary, clear experimental evidence for the existence of two
wobbling bands, based on a positive ($i_{13/2}$) and a negative
($h_{9/2}$) parity configuration, has been observed in $^{183}$Au.
Both the bands are suggested as the transverse wobbling bands
based on the calculations by PRM and HFA models. 
The present work
represents the first observation of TW bands in $A\sim 180$ region and
a first identification of the initial increasing part of wobbling energy as
a function of spin in a TW band, in the whole of the nuclear chart.
One important conclusion from the present work is that the decrease
of $E_\textrm{wob}$ with $I$ may not be considered as a necessary
condition for the evidence of TW. 
Moreover, this is, so far, the only example
of the observation of wobbling bands based on both positive and
negative parity configurations in a single nucleus and represents
the first evidence of triaxial shape coexistence in Au isotopes.

\begin{acknowledgments}
The authors gratefully acknowledge the effort of the cyclotron operators at VECC,
Kolkata for providing a good quality of $^{20}$Ne beam. We thank all the members of
VECC INGA collaboration for setting up the array. 
Partial financial support from Department of Science \& Technology (DST), Govt. of 
India is gratefully acknowledged for the clover detectors of INGA (Grant no No. 
IR/S2/PF-03/2003-II). 
SN, RB and S. Dar acknowledge with thanks the financial support received as 
research fellows from the Department of Atomic Energy (DAE), Govt. of India. 
HP acknowledges the support received from the Ramanujan Fellowship Research Grant 
under SERB-DST Grant No. SB/S2/RJN-031/2016 of Govt. of India. 
The work of Q. B. C. was supported by Deutsche Forschungsgemeinschaft (DFG)
and National Natural Science Foundation of China (NSFC) through funds provided to
the Sino-German CRC 110 ``Symmetries and the Emergence of Structure in QCD" (DFG Grant
No. TRR110 and NSFC Grant No. 11621131001). 
The work of S. F. was supported by the US Department of Energy (Grant No. DE-FG02-95ER40934). 
The work of P.C is supported by the U.S. Department of Energy, Office of Science, Office 
of Nuclear Physics, under Grant No. DE-FG02-94ER40848.
\end{acknowledgments}

\end{document}